\documentclass[12pt,english]{paper}
\usepackage{amsthm,amsmath,amssymb}
\usepackage{mathrsfs}
\usepackage[T1]{fontenc}
\usepackage[latin1]{inputenc}
\usepackage{graphicx}
\usepackage{cite}
\makeatletter

\newcommand{\bibstyle@aas}{\bibpunct{(}{)}{;}{a}{}{,}}

\makeatletter

\makeatletter

\usepackage{geometry}

\makeatletter

\usepackage{epsfig}

\makeatother

\makeatother

\makeatother

\usepackage{babel}
\makeatother
\begin{document}

\title{Direct Measurement of Upward-going Ultrahigh Energy Dark Matter at the Pierre Auger Observatory}

\author{Ye Xu}

\maketitle

\begin{flushleft}
School of Electronic, Electrical Engineering and Physics, Fujian University of Technology, Fuzhou 350118, China
\par
e-mail address: xuy@fjut.edu.cn
\end{flushleft}

\begin{abstract}
In the present paper, it is assumed that there exist two dark matter particles: superheavy dark matter particles (SHDM), whose mass $\sim$ inflaton mass, and light fermion dark matter (DM) particles which are the ultrahigh energy (UHE) products of its decay. The Earth will be taken as a detector to directly search for the UHE DM particles. These upward-going particles, which pass through the Earth and air and interact with nuclei, can be detected by the fluorescence detectors (FD) of the Pierre Auger observatory (Auger), via fluorescent photons due to the development of an EAS. The numbers and fluxes of expected UHE DM particles are evaluated in the incoming energy range between 1 EeV and 1 ZeV with the different lifetimes of decay of SHDM and mass of $Z^{\prime}$. According to the Auger data from 2008 to 2019, the upper limit for UHE DM fluxes is estimated at 90\% C.L. with the FD of Auger. UHE DM particles could be directly detected in the energy range between O(1EeV) and O(10EeV) with the FD of Auger. Thus this might prove whether there exist SHDM particles in the Universe. 
\end{abstract}


\section{Introduction}
It is revealed by cosmology that dark sector dominates in the Universe. Dark matter (DM) comprises of the bulk of matter in the Universe. And many evidences for the existence of DM and its abundance ($26.6\%$) in the Universe are provided by cosmological and astrophysical observations\cite{bergstrom,BHS,Planck2015}. DM is distributed in a halo surrounding a galaxy. One searches for these thermal DM particles via direct and indirect measurements\cite{CDMSII,CDEX,XENON1T,LUX,PANDAX,AMS-02,DAMPE,fermi}. So far no one has found this class of DM particles yet, because of the very small cross section between them and nuclei (maybe O(10$^{-47}$ cm$^2$))\cite{XENON1T,PANDAX}.
\par
The Superheavy Dark Matter (SHDM) $\phi$, based on the possibility of particle production due to time varying gravitational fields in the early Universe, is an alternative DM scenario\cite{KC87,CKR98,CKR99,KT,KCR,CKRT,CCKR,KST,CGIT,FKMY,FKM}. This SHDM has a mass with the same order of magnitude as that of the inflaton. Its lifetime exceeds the age of the Universe $t_0$ ($\sim 10^{17}s$)\cite{AMO,EIP}. Since direct measurement of these SHDM particles or their annihilation products is unattainable, the detection of their decay is an unique and best way to search for them. One has considered the decay of SHDM as a source of high energy cosmic rays in the past\cite{FG,EGS,hill,EGL,BKV,ST,BD,ABK,EIP,BLS,AMO}. That is Standard Model (SM) particles (protons, gamma rays, neutrinos and so on) are produced by the decay of SHDM. In my work, however, it is different from that decay mode that the products of the decay of SHDM are only a ultrahigh energy (UHE) class of light fermion dark matter, not SM particles.
\par
In the present paper, it is made a reasonable assumption that there exist several (at least two) DM species in the Universe. One of these DM species is a non-thermal and non-relativistic SHDM $\phi$ generated by the early universe, which comprises of the bulk of the present-day DM. The other of these DM species is the stable and light DM particles $\chi$ which is UHE products of the decay of SHDM ($\phi\to\chi\bar{\chi}$). The present-day DM may contain a small flux of light DM particles. A Z$^{\prime}$ portal dark matter model\cite{APQ,Hooper} is taken for UHE DM particles $\chi$ to interact with nuclei. And, for the $\chi\chi$Z$^{\prime}$ and qqZ$^{\prime}$ interactions, their vertexes are both assumed to be vector-like. These UHE DM particles could be found by their interaction with nuclei. Thus it is indicated that there exist SHDM particles in the Universe.
\par
The Pierre Auger Observatory (Auger) is a detector array for the measurement of Extreme energy cosmic rays (EECRs), covering an area of 3000 km$^2$, for the measurement of EECRs and located outside the town of Malargue, in the Province of Mendoza, Argentina. EECRs are detected in a hybrid mode at Auger, that is it consists of about 1600 surface detectors (SD) to measure secondary particles at ground level and four fluorescence detectors (FD), each consisting of 6 optical telescopes, to measure the development of extensive air showers (EAS) in the atmosphere\cite{Auger2010}. In the present work, the Earth will be taken as a detector to directly measure the UHE DM particles $\chi$ induced by the decay of SHDM $\phi$ ($\phi\to\chi\bar{\chi}$). These upward-going particles $\chi$, which pass through the Earth and air and interact with nuclei, can be detected by the FD at Auger, via fluorescent photons due to the development of an EAS. In this detection, the main contamination is originated from the diffuse astrophysical neutrinos.
\par
In what follows, the UHE DM and background event rates and UHE DM fluxes will be estimated at the FD of Auger with the different $\tau_{\phi}$. And according to the Auger data, the UHE DM flux limit will be estimated at 90\% C.L. at the FD of Auger in the present paper.

\section{Estimation of UHE DM flux}
The following considers a scenario where the dark matter sector is made up of two particle species in the Universe. One of them is a superheavy dark matter species $\phi$, the other of them is light dark matter species $\chi$ ($m_{\chi} \ll m_{\phi}$), due to the decay of $\phi$, with a very large lifetime. The lifetime for the decay of SHDM to SM particles is strongly constrained ($\tau \geq$ O($10^{26}-10^{29}$)s) by diffuse gamma and neutrino observations\cite{EIP,MB,RKP,KKK}. Then it is considered an assumption that SHDM could only decay to light DM particles $\chi$, not SM particles. So $\tau_{\phi}$ is taken to be between $10^{17}$ s (the age of the Universe) and $10^{26}$ s in the present paper.
\par
The UHE DM flux is made up of galactic and extragalactic components. That is the total flux $\psi_{\chi}=\psi_{\chi}^{Ga}+\psi_{\chi}^{EGa}$, where $\psi_{\chi}{Ga}$ and $\psi_{\chi}^{EGa}$ are the galactic and extragalactic fluxes of UHE DM, respectively. The UHE DM flux from the Galaxy is obtained via the following equation\cite{BLS}:
\begin{center}
\begin{equation}
\psi_{\chi}^{Ga}=\int_{E_{min}}^{E_{max}}F^{Ga}\frac{dN_\chi}{dE_\chi}dE
\end{equation}
\end{center}
with
\par
\begin{center}
\begin{equation}
F^{Ga}=1.7\times10^{-8}\times\frac{10^{26}s}{\tau_{\phi}}\times\frac{1TeV}{m_{\phi}} cm^{-2}s^{-1}sr^{-1}.
\end{equation}
\end{center}
where $\displaystyle\frac{dN_{\chi}}{dE_{\chi}}=2\delta(E_{\chi}-\displaystyle\frac{m_{\phi}}{2})$, and E$_{\chi}$ and N$_{\chi}$ are the energy and number of UHE DM particles, respectively.
\par
The UHE DM flux from the extra galaxy is obtained via the following equation\cite{BGG,EIP}:
\begin{center}
\begin{equation}
\psi_{\chi}^{EGa}=F^{EGa}\int_{E_{min}}^{E_{max}}dE \int_0^{\infty}dz\frac{1}{\sqrt{\Omega_{\Lambda}+\Omega_m(1+z)^3}}\frac{dN_\chi}{dE_\chi}[(1+z)E_\chi]
\end{equation}
\end{center}
with
\par
\begin{center}
\begin{equation}
F^{EGa}=1.4\times10^{-8}\times\frac{10^{26}s}{\tau_{\phi}}\times\frac{1TeV}{m_{\phi}} cm^{-2}s^{-1}sr^{-1}.
\end{equation}
\end{center}
where z denotes the red-shift of the source, $\Omega_{\Lambda}=0.685$ and $\Omega_m=0.315$ from the observation of the PLANCK experiment\cite{Planck2015}. $\displaystyle\frac{dN_{\chi}}{dE_{\chi}}=2\delta(E_{\chi}-\displaystyle\frac{m_{\phi}}{2})$, where E$_{\chi}$ and N$_{\chi}$ denote the energy and number of UHE DM particles, respectively.
\section{UHE DM and neutrino interactions with nuclei}
In the present work, a $Z^{\prime}$ portal dark matter model\cite{APQ,Hooper} is taken for UHE DM particles to interact with nuclei via a neutral current interaction mediated by a new heavy gauge boson $Z^{\prime}$ (see Fig. 1(a) in Ref.\cite{BGG}). Since $\chi\chi Z^{\prime}$ and $qqZ^{\prime}$ are assumed to be vector-like, an effective interaction Lagrangian can be written as follows:
\begin{center}
\begin{equation}
\mathcal{L} = \bar{\chi}g_{\chi\chi Z^{\prime}}\gamma^{\mu}\chi Z^{\prime}_{\mu} + \sum_{q_i} \bar{q_i}g_{qqZ^{\prime}}\gamma^{\mu}q_iZ^{\prime}_{\mu}
\end{equation}
\end{center}
where $q_i$'s are denoting the SM quarks, and $g_{\chi\chi Z^{\prime}}$ and $g_{qqZ^{\prime}}$ are denoting the $Z^{\prime}$-$\chi$ and $Z^{\prime}$-$q_i$ couplings, respectively.
This deep inelastic scattering (DIS) cross section is computed with the fixed parameters (for example, $G=g_{\chi\chi Z^{\prime}}g_{qqZ^{\prime}}$ = 0.05, $m_{\chi}$ = 10 GeV and $m_{Z^{\prime}}$ = 500 GeV) in the lab-frame. This cross section is obtained via the following function(like Fig.1(b) in Ref.\cite{BGG}):
\begin{center}
\begin{equation}
\sigma_{\chi N}=6.13\times10^{-39}\left(\frac{E_{\chi}}{1GeV}\right)^{0.518} cm^2
\end{equation}
\end{center}
where E$_{\chi}$ is the energy of UHE DM particles.
\par
The DIS cross-sction for UHE neutrino interaction with nuclei is computed in the lab-frame and given by simple power-law forms\cite{BHM} for neutrino energies above 1 EeV:
\begin{center}
\begin{equation}
\sigma_{\nu N}(CC)=4.74\times10^{-35}\left(\frac{E_{\nu}}{1 GeV}\right)^{0.251} cm^2
\end{equation}
\end{center}
\begin{center}
\begin{equation}
\sigma_{\nu N}(NC)=1.80\times10^{-35}\left(\frac{E_{\nu}}{1 GeV}\right)^{0.256} cm^2
\end{equation}
\end{center}
where $\sigma_{\nu N}(CC)$ (or $\sigma_{\nu N}(NC)$) is the DIS cross-section for neutrino interaction with nuclei via a charge current (CC) or neutral current (NC). $E_{\nu}$ is the neutrino energy. 
\par
The UHE DM and neutrino interaction lengths can be obtained by
\par
\begin{center}
\begin{equation}
L_{\nu,\chi}=\frac{1}{N_A\rho\sigma_{\nu,\chi N}}
\end{equation}
\end{center}
\par
where $N_A$ is the Avogadro constant, and $\rho$ is the density of matter, which UHE DM particles and neutrinos interact with.
\section{Evaluation of the numbers of UHE DM and neutrinos measured by FD of Auger}
UHE DM particles reach the Earth and pass through the Earth and air, meanwhile these particles interact with matter of the Earth and air. Hadrons are produced by UHE DM interaction with atmospheric nuclei. The secondary particles generated by these UHE hadrons will develop into an EAS. And the most dominant particles in an EAS are electrons moving through atmosphere. Ultraviolet fluorescence photons are emitted by electron interaction with nitrogen. The emitted photons are isotropic and their intensity is proportional to the energy deposited in the atmosphere. A small part of them will be detected by the FD of Auger (see Fig. 1). Since these signatures are similar to DIS of UHE neutrinos, the FD of Auger is unable to discriminate between their signatures. In the present paper, it is made an assumption that there exists air under an altitude of H = 100 km.
\par
The number of UHE DM particles, N$_{det}$, detected by the FD of Auger can be obtained by the following function:
\begin{center}
\begin{equation}
\begin{aligned}
N_{det} & = R\times \int_T \int^{E_{max}}_{E_{min}} \int_{A_{gen}} \int^{\theta_{max}}_0 \eta^{FD}_{trg} \eta^{FD}_{sel}  \Phi_{\chi} P(E,D_e,D)\frac{2\pi {R_e}^2sin(\theta)}{{D_e}^2} d\theta dS dE dt \\
& = R\times \int_T \int^{E_{max}}_{E_{min}} \int_{A_{gen}} \eta^{FD}_{trg} \eta^{FD}_{sel} dS \int^{\theta_{max}}_0 \Phi_{\chi} P(E,D_e,D)\frac{2\pi {R_e}^2sin(\theta)}{{D_e}^2} d\theta dE dt
\end{aligned}
\end{equation}
\end{center}

where $R_e$ is the radius of the Earth and taken to be 6370 km, $\theta$ is the polar angle for the Earth (see Figure 1), $\theta_{max}$ is the maximum of $\theta$ and taken to be $\displaystyle\frac{2\pi}{3}$ for rejecting neutrino events from the spherical crown near the FD of Auger. Here the calculation of the solid angle is simplified by the method of the observational area contraction as a point. R is the duty cycle for Auger and taken to be 15\%\cite{auger_upgrade}. T is the lifetime of taking data for Auger and taken to be 20 years in the present work. dS=dx$\times$dy is the horizontal surface element. E is the energy of an incoming particle and varies from $E_{min}$ to $E_{max}$. Here $\Phi_\chi=\displaystyle\frac{d\psi_\chi}{dE_{\chi}}$. $\eta_{trg}$ and $\eta_{sel}$ are the trigger and selection efficiencies with the FD detector, respectively, and assumed to be related to only the energy for an incoming particle in this evaluation. $S_{eff}=\int_{A_{gen}} \eta_{trg} \eta_{sel} dS$ is the effective area of the FD detector. $A_{gen}$ are the total areas where shower events hit ground level. $S_{eff}$ can be obtained from Ref.\cite{auger2011}:
\begin{center}
\begin{equation}
\begin{aligned}
& \mathcal{A}=\displaystyle\frac{d\mathcal{E}}{dt} =\int_{\Omega} \int_{A_{gen}} \eta^{FD}_{trg} \eta^{FD}_{sel} \eta^{SD}_{trg} \eta^{SD}_{sel} cos(\theta_z) dS d\Omega \\
& = S_{eff}\int_{\Omega} \eta^{SD}_{trg} \eta^{SD}_{sel} cos(\theta_z) d\Omega \\
& S_{eff}=\displaystyle\frac{\mathcal{A}}{\int_{\Omega} \eta^{SD}_{trg} \eta^{SD}_{sel} cos(\theta_z) d\Omega}
\end{aligned}
\end{equation}
\end{center}
where $\mathcal{A}$ is the aperture of the Auger detector. $\theta_z=\displaystyle\frac{\theta}{2}$ is the zenith angle at Auger, and varies from $0^{\circ}$ to $60^{\circ}$. $\Omega$ is the solid angle. $\mathcal{E}$ is the exposure of the Auger detector (see Fig. 11 in Ref.\cite{auger2011}). $\eta^{SD}_{trg}$ and $\eta^{SD}_{sel}$ are the trigger and selection efficiencies with the SD of Auger, respectively. They can be obtained from Ref.\cite{SD2010}.
\par
$P(E,D_e,D)$ is the probability that UHE DM particles interact with air after traveling a distance between $D_e$ and $D_e+D$.
\begin{center}
\begin{equation}
P(E,D_e,D)=exp\left(-\frac{D}{L_{air}}\right)exp\left(-\frac{D_e}{L_{earth}}\right)\left[1-exp\left(-\frac{D}{L_{air}}\right)\right]
\end{equation}
\end{center}
where $D = \displaystyle\frac{H}{cos(\displaystyle\frac{\theta}{2})}$ is the effective length in the detecting zone for the FD of Auger in the air, $D_e=\displaystyle\frac{R_e(1+cos\theta)}{cos\displaystyle\frac{\theta}{2}}$ is the distances through the Earth, and $L_{earth,air}$ are the UHE DM interaction lengths with the Earth and air, respectively.
\par
The diffuse astrophysical neutrinos is roughly estimated with a diffuse neutrino flux of $\Phi_{\nu}=0.9^{+0.30}_{-0.27}\times(E_{\nu}/100 TeV)^{2.13\pm0.13}\times10^{-18} GeV^{-1} cm^{-2}s^{-1}sr^{-1}$\cite{icecube}, where $\Phi_{\nu}$ represents the per-flavor flux, by the above method.

\section{Results}
The numbers of expected UHE DM particles and astrophysics neutrinos are evaluated at different incoming energies (1 EeV < E <1 ZeV) at the FD of Auger, respectively. Since only the deposited energy ($E_{sh}$) in the atmosphere for an EAS can be measured by the FD of Auger, it is important to determine the inelasticity parameter $y$. $y=1 - \displaystyle\frac{E_{\chi^{\prime},lepton}}{E_{in}}$ (where $E_{in}$ is the incoming DM particle or neutrino energy and $E_{\chi^{\prime},lepton}$ is the outgoing DM particles or lepton energy). For UHE DM particles, $E_{sh}=yE_{in}$. Since the EAS due to the neutrino interaction with nuclei via a neutral current is much smaller than that of a charged current, only the charged current be considered in the neutrino interaction with nuclei. Then $E_{sh}=(1-y)E_{in}$ for neutrinos. The mean values of $y$ for UHE DM particles and neutrinos have been computed by Ref.\cite{BGG} and \cite{GQRS}, respectively, and their results are taken to the calculation of $E_{sh}$ in the present paper.
\par
Fig. 2 shows the distributions of expected DM particles and astrophysical neutrinos at different $E_{sh}$'s at the FD of Auger, fixing $m_{Z^{\prime}}$ = 500 GeV. The numbers of expected UHE DM particles can reach about 126 and 1 at the energies with 1 EeV and 13 EeV at the FD of Auger in twenty years with $\tau_{\phi}$ = $10^{19}$ s, respectively, as shown in Fig. 2 (see the red solid line). The ones with $\tau_{\phi}$ = $10^{20}$ s can reach about 13 and 1 at the energies with 1 EeV and 6 EeV at the FD of Auger in twenty years, respectively, as shown in Fig. 2 (see the blue solid line). The ones with $\tau_{\phi}$ = $10^{21}$ s can reach about 1.3 and 1 at the energies with 1 EeV and 1.4 EeV at the FD of Auger in twenty years, respectively, as shown in Fig. 2 (see the purple solid line). The numbers of expected astrophysical neutrinos is at least smaller by 14 orders of magnitude in the energy range between 1 EeV and 100 EeV at the FD of Auger, compared to the ones of expected UHE DMs with $\tau_{\phi}$ = $10^{21}$ s, as shown in Fig. 2 (see the black solid line). So, in this UHE DM detection, the neutrino contamination can be neglected at all. According to the results described above, it is possible that UHE DM particles are directly detected at O(1EeV) at the FD of Auger with $\tau_{\phi} \lesssim 10^{21}$ s.
\par
Fig. 3 shows the numbers of expected DMs and astrophysical neutrinos at the FD of Auger with $m_{Z^{\prime}}$ = 350 GeV, 500 GeV, 1TeV and 2 TeV, respectively, fixing $\tau_{\phi}$ = $10^{19}$ s. The numbers of expected UHE DM particles can reach about 9 and 1 at the energies with 1 EeV and 2 EeV at the FD of Auger in twenty years with $m_{Z^{\prime}}$ = 350 GeV, respectively, as shown in Fig. 4 (see the red solid line). The ones with $m_{Z^{\prime}}$ = 500 GeV can reach about 126 and 1 at the energies with 1 EeV and 13 EeV at the FD of Auger in twenty years, respectively, as shown in Fig. 4 (see the blue solid line). The ones with $m_{Z^{\prime}}$ = 1 TeV can reach about 30 and 1 at the energies with 1 EeV and 88 EeV at the FD of Auger in twenty years, respectively, as shown in Fig. 4 (see the purple solid line). The ones with $m_{Z^{\prime}}$ = 2 TeV can reach about 2 and 1 at the energies with 1 EeV and 9 EeV at the FD of Auger in twenty years, respectively, as shown in Fig. 4 (see the green solid line). UHE DM particles could be directly detected in the energy range between O(1EeV) and O(10EeV) with the FD of Auger with 350 GeV $\lesssim m_{Z^{\prime}} \lesssim$ 2 TeV.
\section{Discussion and conclusion}
For a $Z^{\prime}$ that couples to both leptons and quarks with strengths similar to those of the SM $Z$, its mass is constrained ($m_{Z^{\prime}}\ge$ 2.9 TeV) by the collider observations\cite{BHKN}. For a leptophobic $Z^{\prime}$, however, its mass could be lighter. In the present work, the numbers of expected DM were estimated in the case of the leptophobic $Z^{\prime}$. UHE DM particles cannot be measured at the FD of Auger in the case of the $Z^{\prime}$ that strongly couples to leptons, according to my rough estimation.
\par
Fig. 4 shows the fluxes of expected UHE DM particles with $\tau_{\phi} = 10^{19}$ s (red solid line), $10^{20}$ s (blue solid line) and $10^{21}$ s (purple solid line) and the upper limit for a UHE DM flux at 90\% C.L. (black solid line) at the FD of Auger. According to the Auger data from 2008 to 2019\cite{science,Auger2017,Auger2019}, no upward-going events are measured with the FD of Auger in this period of time. Meanwhile, the expected neutrino events is neglected at all. So the upper limit for the number of UHE DM particles $N_{up}$ is equal to 2.44 at 90\% C.L. with the Feldman-Cousins approach\cite{FC}. This limit excludes UHE DM flux below near 1.5 EeV with $\tau_{\phi}$ = $10^{19}$ s. Thus we know UHE DM particles can be measured at O(1EeV) with the FD of Auger with $\tau_{\phi} \lesssim 10^{21}$ s in the future.
\par
Fig. 5 shows the upper limit for a UHE DM flux at 90\% C.L. and fluxes of expected UHE DM particles with $m_{Z^{\prime}}$ = 350 GeV, 500 GeV, 1 TeV and 2 TeV. Most parameter space cannot be excluded by this upper limit. This need the longer running time for Auger or a detector that is of the larger detecting area.
\par
JEM-EUSO is a fluorescence detector and of the larger observational area (O($10^5$) km$^2$), and its duty cycle $\sim$ 10-20\%\cite{JEM-EUSO}. For measuring UHE DM particles, JEM-EUSO has an advantage over the FD of Auger. So searching for UHE DM particles will depend on the beginning of taking data at JEM-EUSO or the long running time for Auger. This might prove whether there exist SHDM particles in the Universe.
\section{Acknowledgements}
This work was supported by the National Natural Science Foundation
of China (NSFC) under the contract No. 11235006, the Science Fund of
Fujian University of Technology under the contracts No. GY-Z14061 and GY-Z13114 and the Natural Science Foundation of
Fujian Province in China under the contract No. 2015J01577.
\par

\newpage

\begin{figure}
 \centering
 \includegraphics[width=0.9\textwidth]{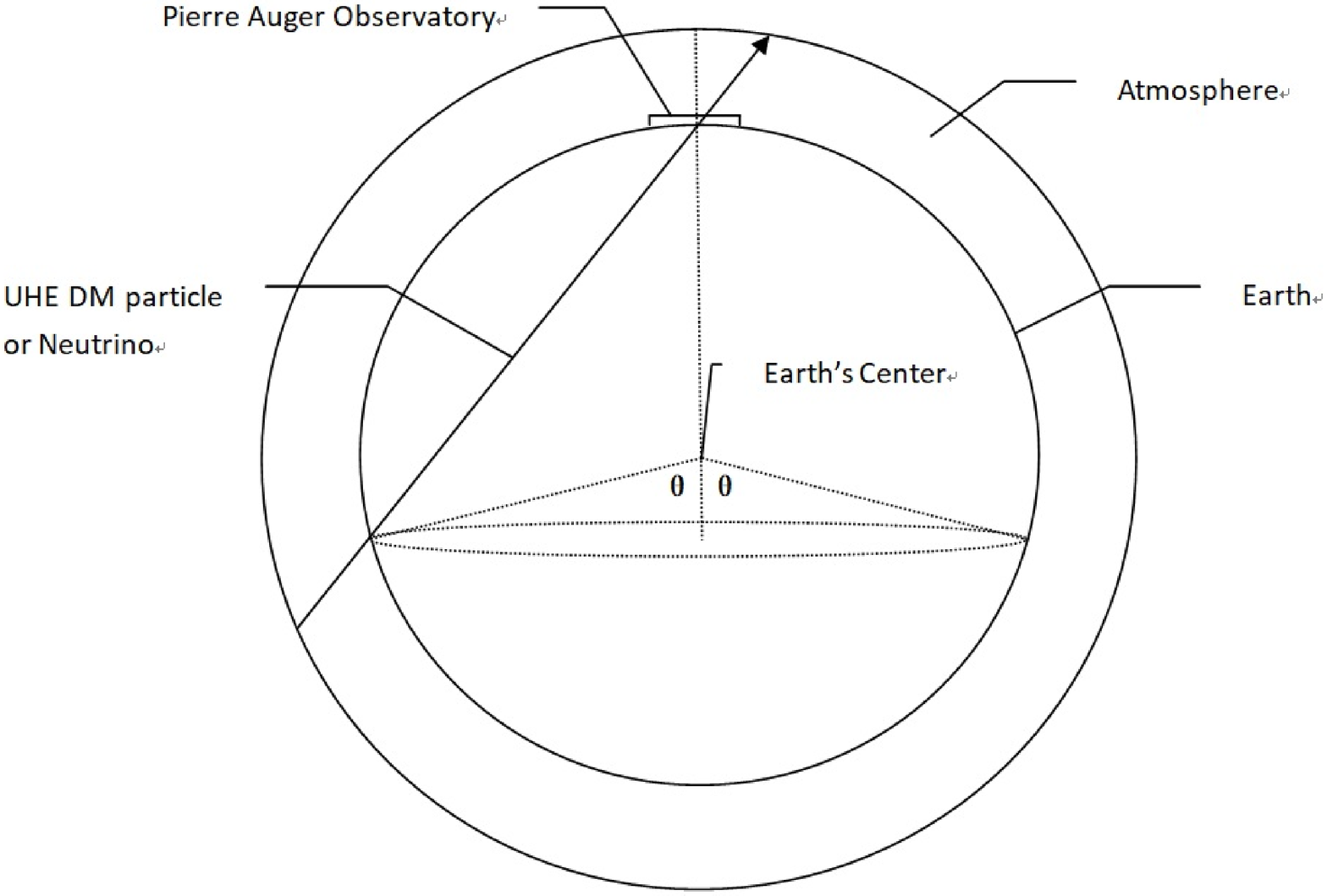}
 \caption{UHE DM particles pass through the Earth and air and can be measured by the FD of Auger, via fluorescent photons due to the development of an EAS. $\theta$ is the polar angle for the Earth.}
 \label{fig:figure_1}
\end{figure}
\newpage
\begin{figure}
 \centering
 \includegraphics[width=0.9\textwidth]{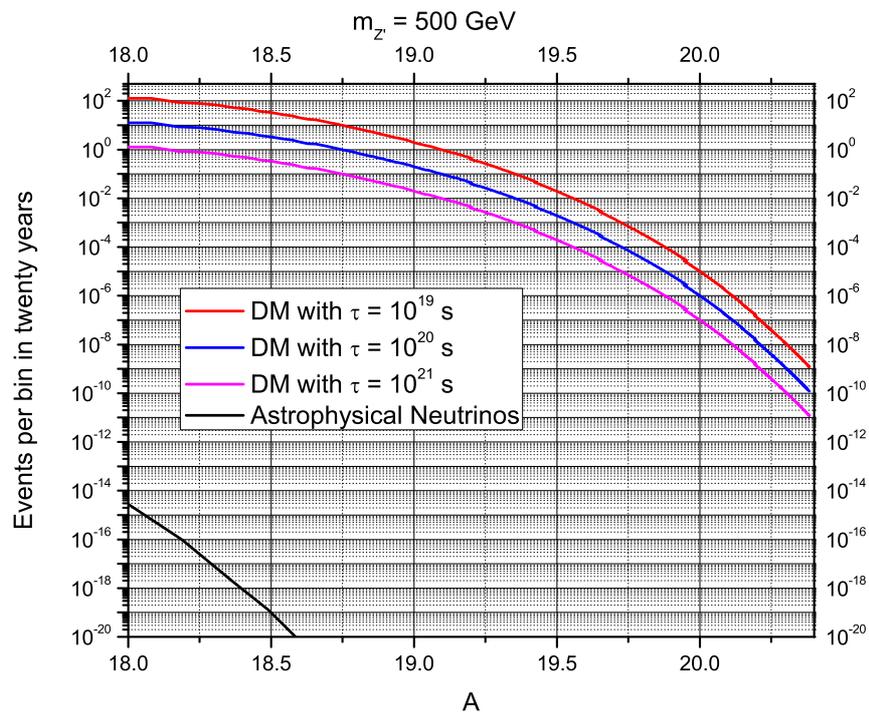}
 \caption{The distribution of expected UHE DM and astrophysical neutrino event rates at the FD of Auger. The red solid line is denoting the numbers of expected UHE DM particles with $\tau_{\phi}=10^{19}$ s. The blue solid line is denoting the ones with $\tau_{\phi}=10^{20}$ s. The blue solid line is denoting the ones with $\tau_{\phi}=10^{21}$ s. The black solid line is denoting the numbers of expected astrophysical neutrinos.}
 \label{fig:event_tau}
\end{figure}

\newpage
\begin{figure}
 \centering
 \includegraphics[width=0.9\textwidth]{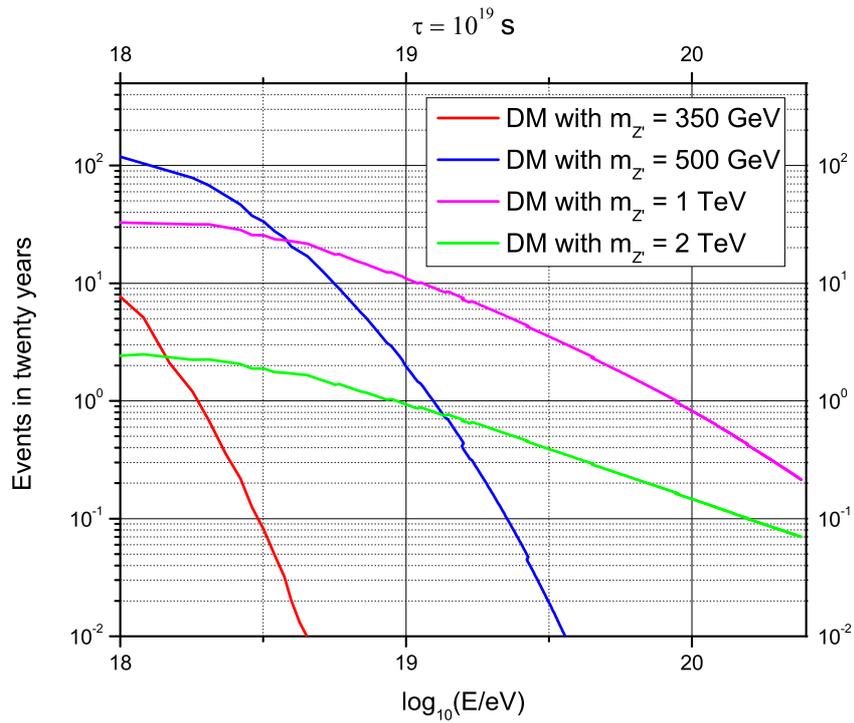}
 \caption{The numbers of expected UHE DM particles and astrophysical neutrinos at the FD of Auger. The red solid line is denoting the numbers of expected UHE DM particles with $m_{Z^{\prime}}$ = 350 GeV. The blue line is denoting the ones with $m_{Z^{\prime}}$ = 500 GeV. The purple solid line is denoting the ones with $m_{Z^{\prime}}$ = 1 TeV. The green solid line is denoting the ones with $m_{Z^{\prime}}$ = 2 TeV.}
 \label{fig:event_mz}
\end{figure}
\newpage
\newpage
\begin{figure}
 \centering
 \includegraphics[width=0.9\textwidth]{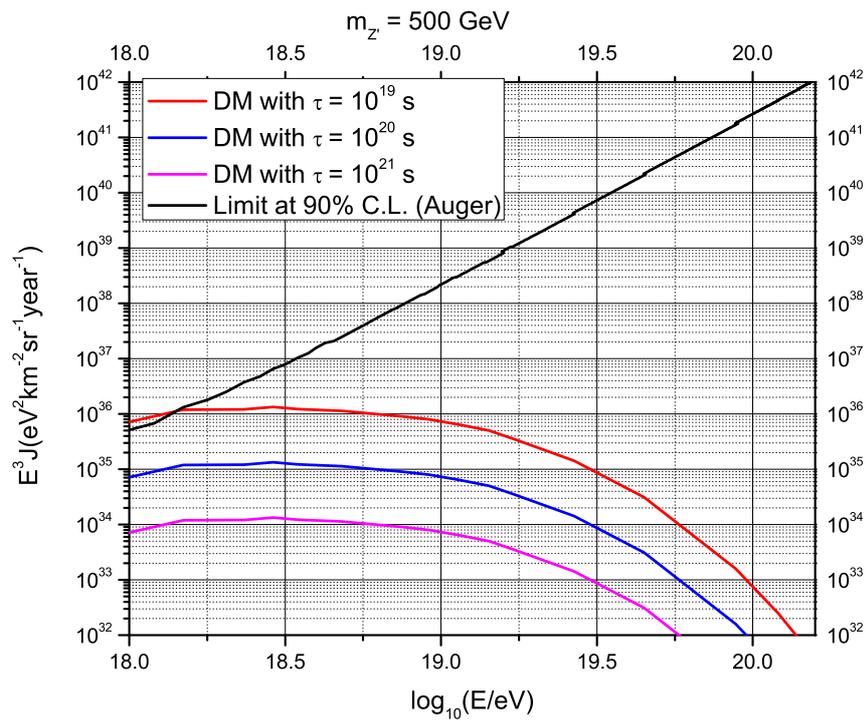}
 \caption{The fluxes of expected UHE DM particles and their upper limit at 90\% C.L. at the FD of Auger. The red solid line is denoting the flux of expected UHE DM particles with $\tau_{\phi}$ = $10^{19}$ s. The blue solid line is denoting the one with $\tau_{\phi}$ = $10^{20}$ s. The purple solid line is denoting the one with $\tau_{\phi}$ = $10^{21}$ s. The black solid line is denoting their upper limit at 90\% C.L. at the FD of Auger.}
 \label{fig:flux_tau}
\end{figure}
\begin{figure}
 \centering
 \includegraphics[width=0.9\textwidth]{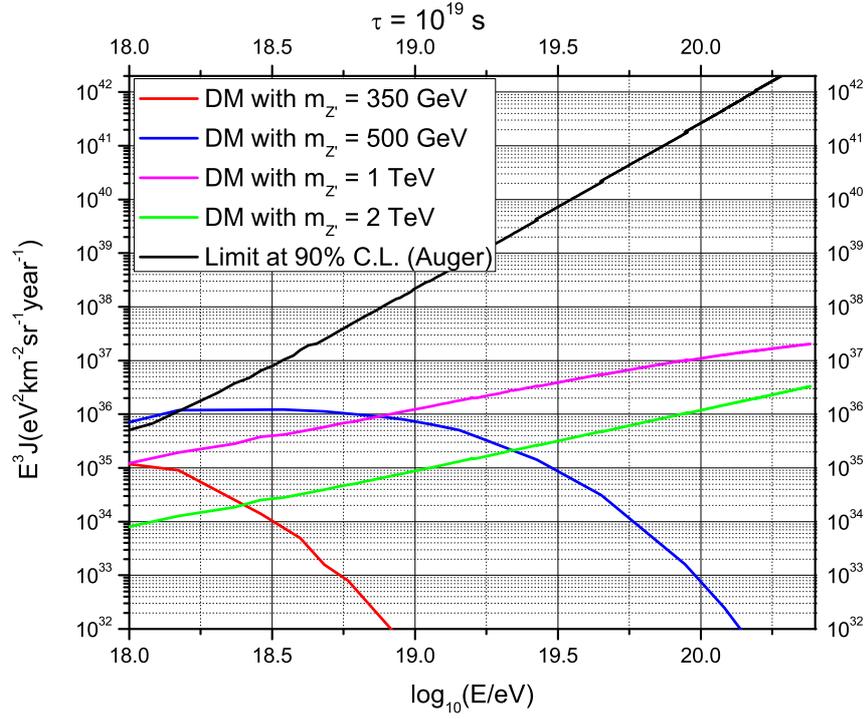}
 \caption{The fluxes of expected UHE DM particles and their upper limit at 90\% C.L. at the FD of Auger. The red solid line is denoting the flux of expected UHE DM particles with $m_{Z^{\prime}}$ = 350 GeV. The blue solid line is denoting the one with $m_{Z^{\prime}}$ = 500 GeV. The purple solid line is denoting the one with $m_{Z^{\prime}}$ = 1 TeV. The green solid line is denoting the one with $m_{Z^{\prime}}$ = 2 TeV. The black solid line is denoting their upper limit at 90\% C.L. at the FD of Auger.}
 \label{fig:flux_mz}
\end{figure}

\end{document}